\documentclass[%
 twocolumn,
showpacs,
 amsmath,amssymb,
 aps,
prb,
]{revtex4-1}

\makeatletter
\def\@xfootnote[#1]{%
  \protected@xdef\@thefnmark{#1}%
  \@footnotemark\@footnotetext}
\makeatother

\usepackage[percent]{overpic}
\usepackage{wrapfig}
\usepackage{xcolor}
\usepackage{lipsum}
\usepackage{graphicx}
\usepackage{epstopdf}
\usepackage{dcolumn}
\usepackage{bm}
\usepackage{bigints}

\begin{document}


\title{Quantifying Confidence in Density Functional Theory Predicted Magnetic Ground States}

\author{Gregory Houchins}
\affiliation{Department of Physics, Carnegie Mellon University, Pittsburgh, Pennsylvania 15213, USA}
\author{Venkatasubramanian Viswanathan}%
\email{venkvis@cmu.edu}
\affiliation{Department of Physics, Carnegie Mellon University, Pittsburgh, Pennsylvania 15213, USA}%
\affiliation{Department of Mechanical Engineering, Carnegie Mellon University, Pittsburgh, Pennsylvania 15213, USA}
\date{\today}

\begin{abstract}
Density functional theory (DFT) simulations, at the generalized gradient approximation (GGA) level, are being routinely used for material discovery based on high-throughput descriptor-based searches.  The success of descriptor-based material design relies on eliminating bad candidates and keeping good candidates for further investigation. While DFT has been widely successfully for the former, often times good candidates are lost due to the uncertainty associated with the DFT-predicted material properties.  Uncertainty associated with DFT predictions has gained prominence and has led to the development of exchange correlation functionals that have built-in error estimation capability.  In this work, we demonstrate the use of built-in error estimation capabilities within the BEEF-vdW exchange correlation functional for quantifying the uncertainty associated with the magnetic ground state of solids.  We demonstrate this approach by calculating the uncertainty estimate for the energy difference between the different magnetic states of solids and compare them against a range of GGA exchange correlation functionals as is done in many first principles calculations of materials. We show that this estimate reasonably bounds the range of values obtained with the different GGA functionals. The estimate is determined as a post-processing step and thus provides a computationally robust and systematic approach to estimating uncertainty associated with predictions of magnetic ground states. We define a confidence value (c-value) that incorporates all calculated magnetic states in order to quantify the concurrence of the prediction at the GGA level and argue that predictions of magnetic ground states from GGA level DFT is incomplete without an accompanying c-value. We demonstrate the utility of this method using a case study of Li and Na-ion cathode materials and the c-value metric correctly identifies that GGA level DFT will have low predictability for $\mathrm{NaFePO_4F}$.  Further, there needs to be a systematic test of a collection of plausible magnetic states, especially in identifying anti-ferromagnetic (AFM) ground states. We believe that our approach of estimating uncertainty can be readily incorporated into all high-throughput computational material discovery efforts and this will lead to a dramatic increase in the likelihood of finding good candidate materials.

\end{abstract}

\maketitle


\section{\label{sec:level1}Introduction}
There is an explosion in the computation of material properties based on off-the-shelf density functional theory software and this has led to a rapid acceleration in material discovery in a variety of areas such as energy\cite{energyDFT} and biology.\cite{bioDFT}  Increasingly, the results of these calculations are driving material design choices.\cite{Curtarolo2013,holdren2014materials} There are numerous success stories of computationally-driven material discovery largely based on high throughput computation of one or more descriptors, in thermoelectrics \cite{Zhang2016}, electrocatalysis\cite{Greeley2006}, battery materials \cite{Kang2006}, hydrogen storage \cite{Wolverton2008}, topological insulators \cite{Yang2012}, and magnetic materials. \cite{Debrov2013}

The challenge of using high-throughput materials discovery is that typically large searches lead to only a few candidates. One example is a high-throughput search of over 5400 oxide/oxynitride compounds for solar light capture leading to only 15 new candidates.\cite{Castelli} It is now widely acknowledged that high-throughput discovery based on density functional theory calculations, largely at the generalized gradient approximation (GGA) level,\cite{GGA} is excellent at eliminating bad candidates but not as good in spotting and keeping the good candidates.  An emerging frontier is to incorporate uncertainty in order to improve the predictability and aid in high-throughput discovery as understanding materials with high uncertainty within a model may lead to the discovory of new phenomena.\cite{Citrine} 

One of the most useful yet potentially troubling aspect of DFT is the easy accessibility of meta-stables states.\cite{Sun2016} In the context of magnetic materials, this means a whole collection of magnetic states can be attained for the same material and structure. This makes predicting the true magnetic ground state very challenging.\cite{MagCathode1,MagCathode2} Conventionally, only a single energy difference between two states could be attained when using any one exchange-correlation functional. One naive way to increase the likelihood of an accurate prediction is to use multiple exchange correlation functionals.  However, the selection of these functionals is not systematic and unlikely to guarantee realistic uncertainty bounds associated with the ground state prediction and in the case where GGA level DFT is insufficient to capture the correct properties the few functionals tested may give inconsistent results. The Bayesian error estimation functional (BEEF)\cite{BEEF} carries with it a prediction of uncertainty calibrated to recreate the error as mapped to experimental training data by generating a collection of GGA level functionals. This empirical error estimation has been used to quantify uncertainty in heterogenous catalysis,\cite{MedfordScience} electrocatalysis \cite{SidACS-Catalysis} and mechanical properties for solid electrolytes.\cite{ZeeshanPRB} In this work, we extend the framework of uncertainty estimation and demonstrate a robust approach for quantifying uncertainty of the ground states of magnetic materials.  Magnetic materials are important in a wide variety of applications such as information storage \cite{magnetic_information}, colossal magnetoresistance\cite{CMR}, spin currents \cite{spin_current}, and medicine. \cite{medicine} In our approach, we calculated an ensemble of energy difference between different magnetic states (for e.g., ferromagnetic and anti-ferromagnetic) of the same material.  From that ensemble, we can define a confidence value (c-value) that quantifies the agreement between GGA level functionals and ultimately the certainty that one spin state is more energetically favorable than the other. Our approach is computationally efficient, as compared to doing many different calculations based on different functionals, simulating non-self consistently thousands of functionals using one self consistent calculation. We demonstrate the utility of the developed method and the c-value with a case study of Li-ion and Na-ion cathode materials. We propose that such a method needs to be used to ensure a sufficient level of agreement for the magnetic ground state, and thus, the derived properties such as voltage, electronic conductivity of the magnetic material. 

\section{Methods}
\subsection{Ground state prediction}

The magnetic ground state is defined as the spin configuration that minimizes the energy. 
\[\mathrm{E_{GS}}=\textrm{min[E(}\bm{S_1,S_2,\cdots, S_i} )] \]

Naively, all other parameters other than spin configuration should be fixed and only the spin is changed. In the context of DFT, the exact cell parameters and atoms locations that give the lowest energy may vary for different initial spin states. It is therefore best give a create a collection of initial spin states that are to be compared and for each spin state, allow the lattice parameters and atomic positions to vary. 

Another issue in the prediction of magnetic states is the ambiguity of "antiferromagnetic" (AFM) for materials with certain geometries or  with multiple distances of magnetic interactions that could lead to different magnetic couplings. Layered honeycomb materials provide a great example of a geometry that leads to ambiguity as they can have a zigzag antiferromagnetic structure (Type I) such as that seen in $\mathrm{FePSe_3}$ or where all three nearest-neighbors are antiferromagnetically coupled (Type II) as seen in $\mathrm{MnPSe_3}$.\cite{layered_magnet} Perovskites on the other hand demonstrate the ambiguity of various length scales. Atoms can be inter-plane AFM and intra-plane FM (A-Type) as in $\mathrm{LaMnO_3}$ at low temperature \cite{LaMnO3_lowT}, intra-plane AFM and inter-plane FM (C-Type)as in $\mathrm{CaCrO_3}$ and $\mathrm{SrCrO_3}$ \cite{CaCrO3,ACrO3,SrCrO3}, or both intra-plane and inter-plane AFM (G-Type) as in $\mathrm{LaTiO_3}$ \cite{LaTiO3}.  Studies of magnetic states compare typically two (or a few) states: the obvious ferromagnetic (FM) where all spin point in the same direction and one of the possibly many antiferromagnetic orientations. For example, a first-principles study by Baetting \textit{et. al.}\cite{Baettig2005} of perovskite multiferics calculated the energy difference for ferromagnetic and G-type like ferrimagnet (FiM), but did not study the A-type or C-type couplings. It is worth pointing out that the difference in energy between these two states does not necessarily predict the global ground state but rather which of the two is more favorable. In materials with a variety of magnetic range interactions, it is conceivable that other AFM states could exist. This can be seen again in the context of perovskites that can have intra-plane AFM, inter-plane AFM, or both. It should therefore be understood that in some cases, several spin configurations must be tested for an accurate prediction. 

\subsection{Bayesian Error Estimation}
Recently, Baysian Ensemble Error Functional with van der Waals correlations (BEEF-vdW)\cite{BEEF} has provided a way to systematically estimate the uncertainty of a DFT calculation. This empirically fit functional is a generated from a generalized gradiant approximation (GGA) exchange energy and Perdew-Burke-Ernzerhof (PBE), Perdew-Wang68 local density approximation (LDA), and vdW–DF2 non-local (nl) correlation contributions. The exchange enhancement factor, is fit using an expansion in terms of Legendre Polynomials $B_m$, given by
\begin{align}
F_x^{GGA}(s)&=\sum_{m}a_m B_m[t(s)],\\
t(s)&=\dfrac{2s^2}{4+s^2}-1,~ -1\leq t \leq 1.
\end{align}
 Therefore, the exchange-correlation energy is given by
\begin{multline}\label{eq:1}
E_{xc}=\sum_{m}\int \epsilon_x^{UEG}(n)B_m[t(s)]d\bm{r} 
+\alpha_c E^{LDA-c}\\+(1-\alpha_c)E^{PBE-c}+E^{nl-c}
\end{multline}

The parameters of the functional $a_m$ and $\alpha_c$ are optimized with respect to a collection of experimental data. To generate an ensemble of functionals a distribution of these parameters is generated. Therefore, once a self consistent DFT calculation has been performed using the best fit parameters, the converged electron density $n(\bm{r})$ can be used along with the spread of $\alpha_c$ and $a_m$ to generate an ensemble of energies non-self-consistently using Equation (\ref{eq:1}).  The spread of these values is tuned to create a spread in energies that recreates the error of the best fit DFT calculation with respect to the experimental training data. The data sets include molecular formation energies and reaction energies, molecular reaction barriers, noncovalent interactions, solid state properties such as cohesive energies and lattice constants, and chemisorption on solid surfaces.  In this way, the error estimation has been trained to predict how uncertain the prediction is with respect to known uncertainties in experiment. 

\section{Benchmarking}
It is worth highlighing that the BEEF-vdW functional was trained on data sets that did not explicitly include any magnetic properties in the training sets. We can benchmark the accuracy of DFT calculations using the BEEF-vdW functional in capturing the magnetic properties of materials by comparing the calculated magnetization to experiment measurements and other GGA level functionals. The calculation of atomic magnetic moments is carried out in the trivial way by integrating the difference of spin up and spin down electron densities converged from a spin polarized calculation.\cite{spin-polarised} In this way, the predicted magnetic moment is a z-projection of the spin-only magnetic moment, neglecting orbital magnetization. The approximation of spin only magnetization can be made since the the orbital moment in the case of transition metals is quenched from the delocalization and band formation of electrons in the bulk. The calculation of the z-projection of spin only can be justified through the ability to derive a Stoner model through this formalism,\cite{Stoner1,Stoner2,Stoner3,Stoner4} as well as this method is in agreement with covalent description of magnetism. \cite{covalent_magnetism} We therefore set the initial guess for magnetic moments for each atom based on a spin only estimation that depends only on the number of unpaired electrons, $n$, \[M=g\mu_BS_z=\mu_B n,   \] where the gyromagnetic ratio, $g$ is two and the z-component of spin for each unpaired electron is $\frac{1}{2}$.    

We find the BEEF-vdW functional predicts the magnetic moments with similar accuracy to other GGA functionals. Table \ref{tab:2} shows the various GGA functional predictions of magnetic moments for various materials using the experimental lattice parameters. The overestimation in magnetic moment for bulk Cr is well known in DFT due to the fact that the experimental ground state is an incommensurate spin density wave \cite{DFT-Cr,Chromium}, while the error in $\mathrm{CuCr_2O_4}$ compared to experiments is due to the noncolinear structure of the ground state.\cite{CuCr2O4} The DFT calculation converged to the collinear version of the experimentally seen magnetic configuration, properly capturing the total magnitude of the spins rather than the z-projection.

\begin{table}
\caption{\label{tab:2} The magnetic moment per magnetic ion in $\mu_B$ as predicted by BEEF-vdW functional in the first row compared to experimental measurements in the second. In the case of $\mathrm{CuCr_2O_4}$ the total magnetic moment is given.}
\begin{ruledtabular}
\begin{tabular}{lccccccc}

     &Fe & Cr & Ni & \begin{tabular}{c} $\mathrm{FePO_4}$-q\\$\mathrm{FePO_4}$-o\end{tabular}& $\mathrm{LaMnO_3}$ & $\mathrm{CuCr_2O_4}$ \\
\\ \hline
\\
BEEF&
2.33&
1.62&
0.61&
\begin{tabular}{l}
4.29\\
4.03
\end{tabular} &  
3.89& 
5.00
\\
\\

PBE&
2.13 &
1.23 &
0.60 &
\begin{tabular}{l}
4.31\\
4.00
\end{tabular} &  
3.85 & 
5.00 
\\
\\

RPBE&
2.21 &
1.77 &
0.61 &
\begin{tabular}{l}
4.33\\
4.02
\end{tabular} &  
3.89 & 
5.00 
\\
\\

PBEsol&
2.01 &
0.74 &
0.58 &
\begin{tabular}{l}
4.30\\ 
3.97
\end{tabular}&
3.77 & 
5.00 
\\
\\

Expt.&
2.22\footnotemark[1]&
0.62\footnotemark[2]&
0.61\footnotemark[1]& 
\begin{tabular}{l}
4.53\footnotemark[3] \\ 
4.02\footnotemark[4] 
\end{tabular}&
3.70\footnotemark[5]& 
0.39\footnotemark[1]
\end{tabular}
\end{ruledtabular}
\footnotetext[1]{Ref.~\citenum{AIP_handbook}}
\footnotetext[2]{Ref.~\citenum{Chromium}}
\footnotetext[3]{Ref.~\citenum{FePO4_quartz}}
\footnotetext[4]{Ref.~\citenum{FePO4_LiFePO4_magmom}}
\footnotetext[5]{Ref.~\citenum{LaMO3_high_T}}
\end{table}

\section{Prediction of magnetic ordering}
In the case of new materials without a known spin structure, DFT can be used to predict the most energetically favorable configuration. We demonstrate this search for the correct magnetic state with previously characterized materials to demonstrate our method. Starting from both ferromagnetic and antiferromagnetic spin states, we optimize the lattice constants by starting from an experimentally derived unit cell and minimizing the energy with respect to volume by scaling all lattice constants uniformly. At lease five different volumes were tested with the volumes varying with a strain parameter with respect to the experimental cell of $x=V/V_0=0.9,0.95,1.0,1.05,1.10$ in most cases and $x=0.95,0.975,1.0,1.025,1.05$ in others where convergence was an issue. Each cell volume and its corresponding energy, converged to $<0.1$ meV, is fit with a third order polynomial stabilized jellium equation of state (SJEOS) \cite{eos} to find the strain parameter corresponding the minimum energy. Once the cell paramters are identified, the internal coordinates of the atoms are allowed to relax to a maximum force of less than 0.01 ev/\AA{}. All calculations are performed using the BEEF-vdW functional in GPAW, a real space grid implementation of the projector augmented-wave method \cite{GPAW}, with a $8\times 8\times 8$ k-point mesh and a real space grid spacing of $h=0.18 ~\mathrm{\AA{}}$. After the cell and atomic coordinates are optimized, the converged electron density for this configuration is used to generate an ensemble of 2000 energy values non-self consistently. Doing this for two magnetic states and subtracting the ensemble of energies element-wise provides an ensemble of energy difference for error estimation. The results of our calculations, where we have compared FM and the lowest lying AFM state, are shown in Table \ref{tab:1}. For comparison, the same optimization procedure was repeated for three other functionals at the GGA level: Perdew-–Burke--Ernzerhof (PBE)\cite{GGAPBE}, Revised Perdew-–Burke–-Ernzerhof (RPBE)\cite{GGARPBE} and PBEsol\cite{GGAPBEsol}. 

The materials used were chosen to represent a range of crystal structures, complexity, and elements, as well as different mechanisms of magnetism. Lattice distortions play a large role in the case of $\mathrm{LaMnO_3}$,\cite{sawada1997} and $\mathrm{CuCr2O4}$\cite{CuCrO4JT}. Direct exchange accounts for the ferromagnetic nature of Fe and Ni,\cite{van1945} while indirect exchange acounts for the AFM tendencies of the oxides $\mathrm{FePO_4,~LaMnO_3,~and~CuCr_2O_4}$. 

The optimization procedure involves at least ten self-consistent calculations just for cell optimizations. To test the possibility of predicting the magnetic ground state without optimizing the lattice constants to save computation time, we performed only an internal geometry relaxation on the experimental cell parameters. The results of this are in Table \ref{tab:3}. The two methods of optimizing and using the experimental lattice constant give very similar results as seen in Table \ref{tab:4}

\begin{figure}[]
\begin{minipage}[t]{0.4\linewidth}
  \begin{overpic}[width=1.05\linewidth,trim={0 2cm 0 2cm},clip]{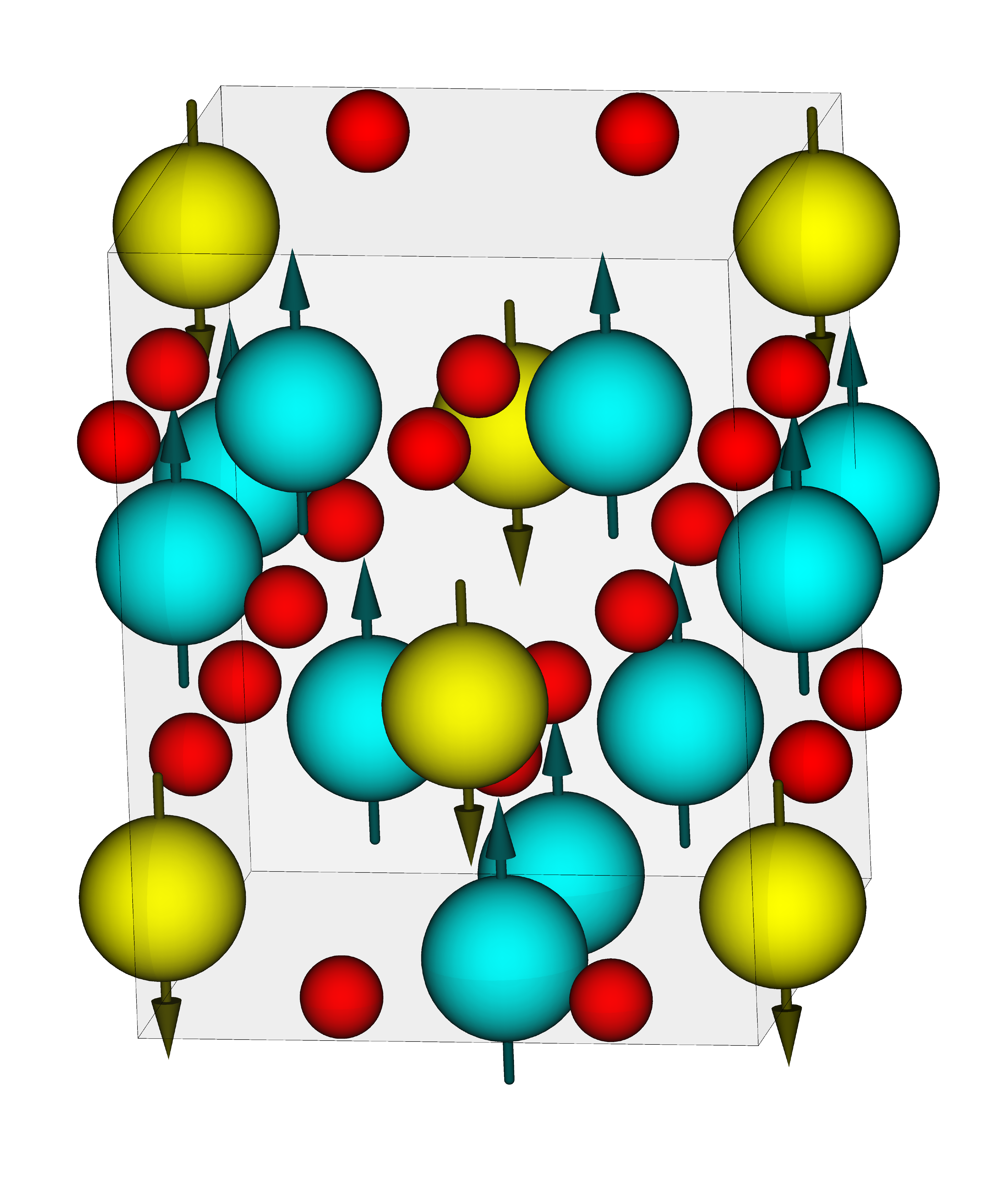}
   \put (30,83) {\textbf{AFM1}}
  \end{overpic}
    \begin{overpic}[width=1.05\linewidth,trim={0 2cm 0 2cm},clip]{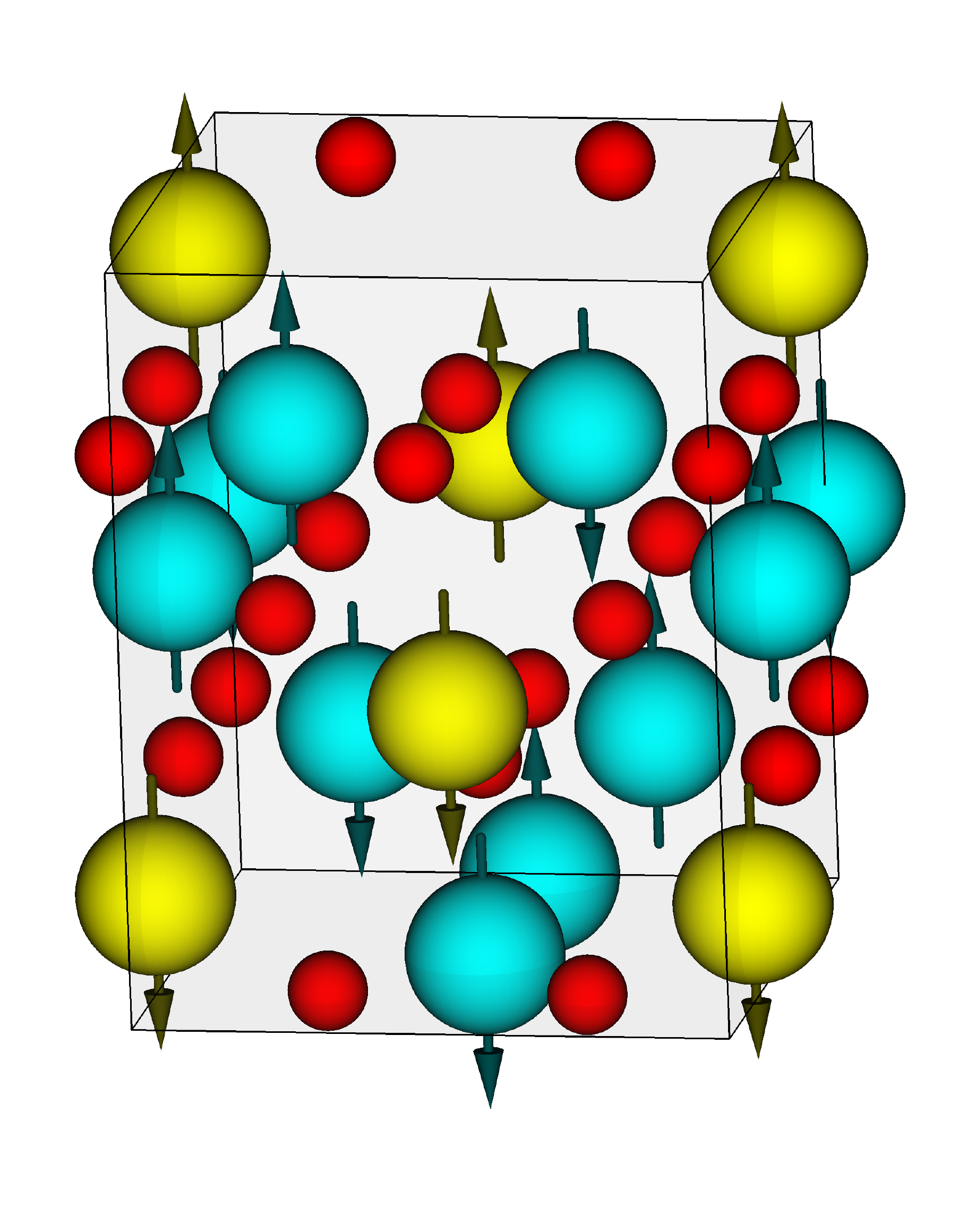}
     \put (30,85) {\textbf{AFM2}}
    \end{overpic}
\end{minipage}
\begin{minipage}[]{0.58\linewidth}
  \includegraphics[width=1.0\linewidth,trim={.5cm 0 1.5cm 0},clip]{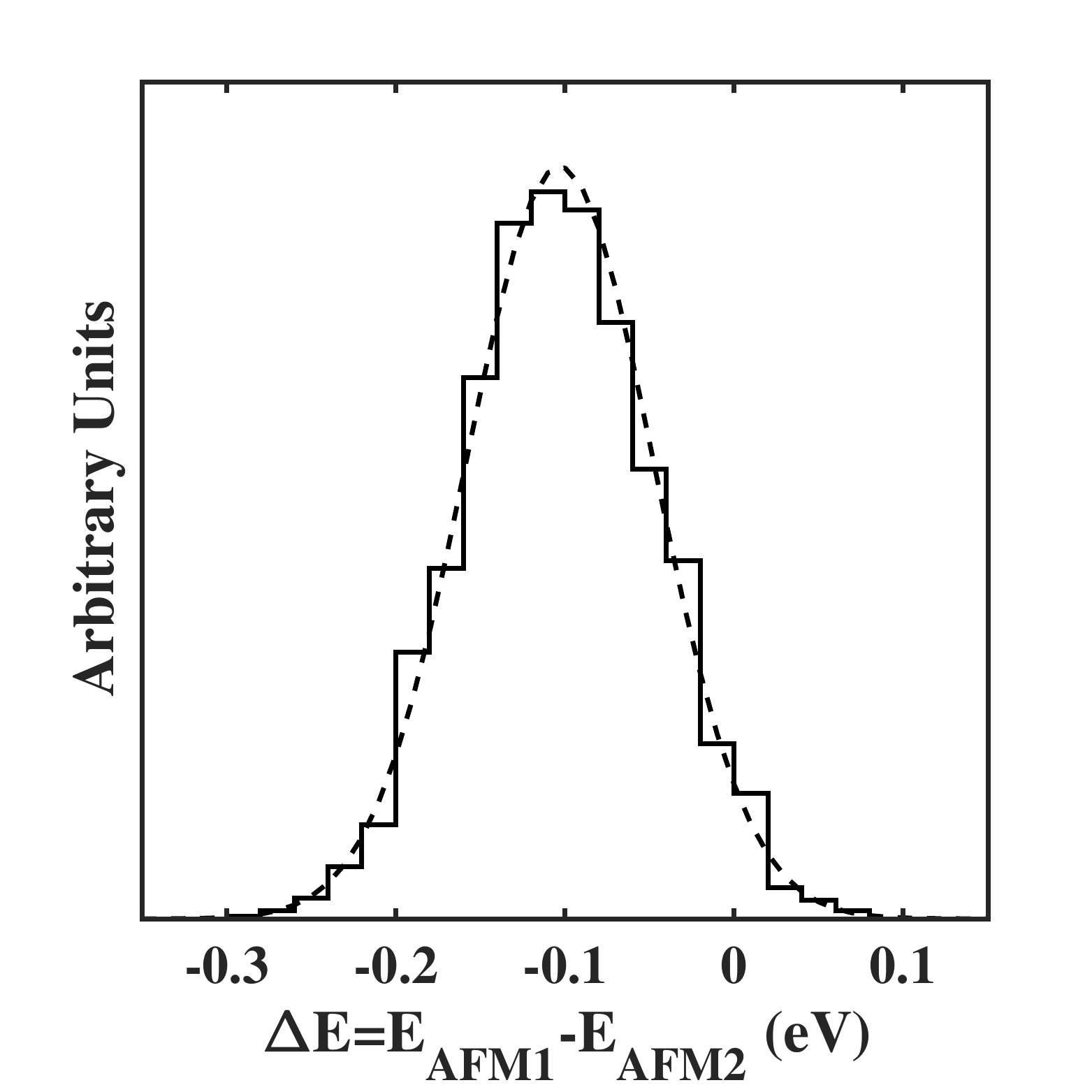}
\end{minipage}

\caption{\label{fig:CuCr2O4} The two converged magnetic states and the distribution of energy differences. For this structure all cases with initially ferromagnetic spin structure converge to one of these two antiferromagnetic cases.}
\end{figure}

In most cases, the lowest lying states were simple ferromagnetic and anti-ferromagnetic configurations. In the case of $\mathrm{CuCr_2O_4}$, however, the antiferromagnetic coupling between copper and chromium prevented any DFT calculation from converging to completely ferromagnetic state. The two antiferromagnetically aligned states are seen in Figure \ref{fig:CuCr2O4}.

\section{Prediction confidence}
In order to measure the confidence of the prediction of the magnetic ground state, we define a c-value as the percentage of the ensembles that support the hypothesis of the best fit functional. For example, in the case of a predicted ferromagnetic state this would be
\begin{equation}
c =  \frac{1}{N_{ens}} \sum_i^{N_{ens}} \prod_{j} \Theta(E_{AFM_j,i}-E_{FM,i})
\end{equation}
where N$_{ens}$ is the number of functionals used, the sum is over functionals, the product is over all magnetic states other than the predicted state, and $\Theta(x)$ is the Heaviside step function. 
The method for calculating the c-value requires the entire dataset of energy differences for all of the possible magnetic states and therefore cannot be easily recalculated or utilized for future independent studies. We therefore can approximate the c-value by modeling the spread of energy differences between the predicted ground state and other possible magnetic state as a normal distribution with ensemble mean $\mu$ and calculated standard deviation $\sigma$. We can then integrate the normalized distribution as an approximation of the Heaviside function. That is for a material that we have again predicted to be ferromagnetic, the approximate c-value is 
\begin{equation}
c \approx   \prod_{j} \int_{0}^{\infty} \frac{\mathrm{d}x}{\sqrt{2\pi\sigma_j^2}}e^{-\frac{(x-\mu_j)^2}{2\sigma_j^2}}
\end{equation}
Again, the product is over all magnetic states other than the predicted state. 

\begin{figure}
\includegraphics[width=.5\textwidth]{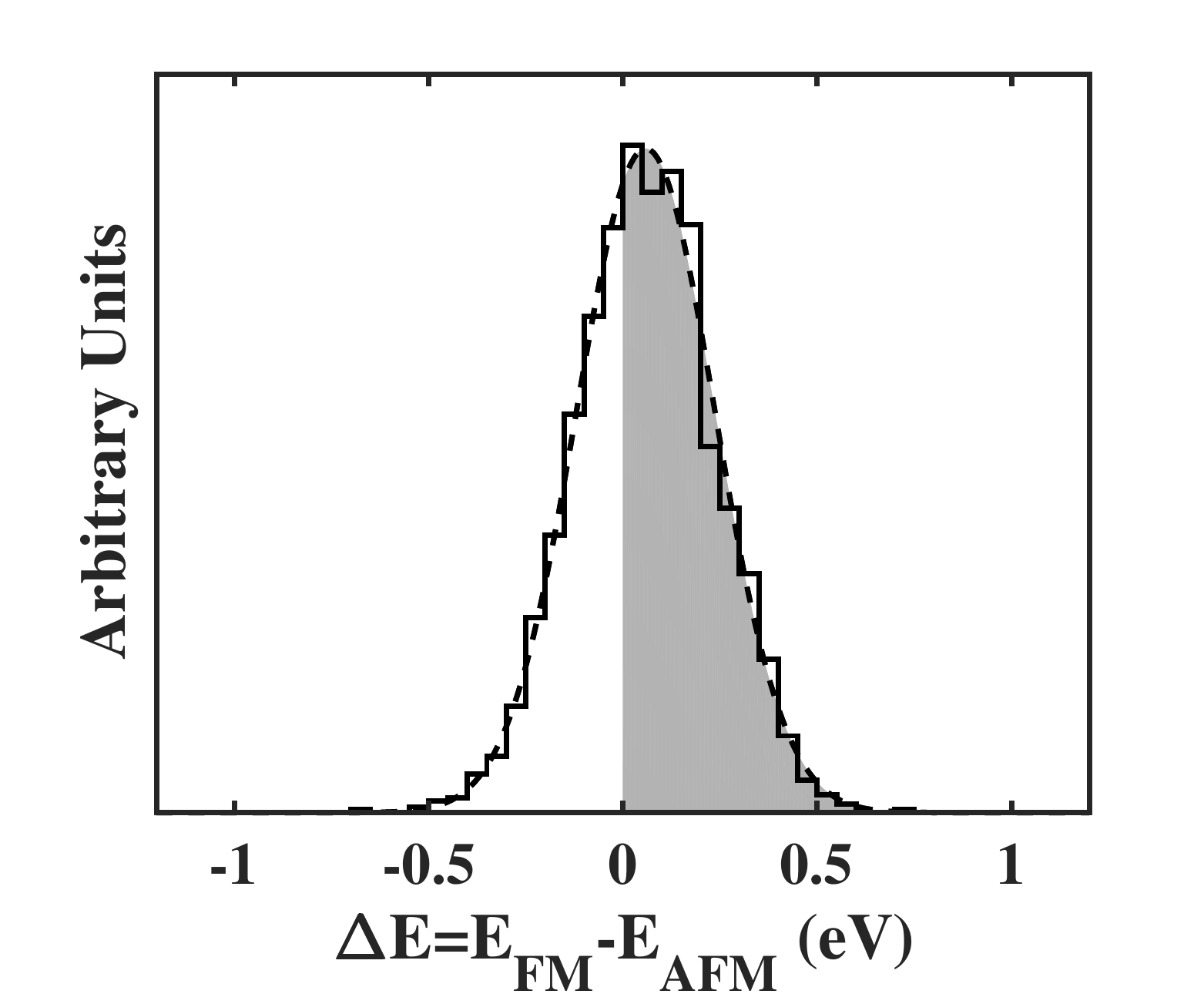}
\caption{\label{Cr} A histogram of the n=2000 ensemble of energy differences with 50 bins. The area under the histogram to the right of 0 represents the c-value and the area of the gray shaded region represents the appoximate c-value. The normal distribution guess created from the calculated mean and standard deviation is also shown in good agreement with raw data. }
\end{figure}

The two methods of calculating prediction c-value give results consistent to within .01 and therefore can be used relatively interchangeably. The approximate c-value is expected to get worse as the number of tested magnetic states increases. Although the method of counting the exact number of functionals in agreement is the most accurate, it requires access to the raw data. The approximation to a normal distribution, however, requires only the mean and standard deviation for each energy difference between all magnetic states tested and has negligible deviation from the exact calculated confidence. A depiction of this c-value as well as the approximate c-value in the case of only two magnetic states tested can be seen in Figure \ref{Cr}. 

\begin{table*}
\caption{\label{tab:1}Calculated magnetic energy difference for optimized structures $E_{FM}-E_{AFM}$, unless otherwise noted, of various crystal structures is given so that the prediction of BEEF-vdW can be compared to that of other GGA-level functionals. The BEEF-vdW energy difference is accompanied by the ensemble standard deviation. The mean of the ensemble of energy difference generated by BEEF-vdW, c-value incorperating all magnetic states, as well as the approximated c-value also shown for each material. All energies are in eV}
\begin{ruledtabular}
\begin{tabular}{lcccccccc}

Material & BEEF-vdW  & PBE & RPBE & PBEsol & exp & \begin{tabular}{c} Ensemble\\ Mean \end{tabular} & c-value & \begin{tabular}{c} Approx.\\ c-value \end{tabular}
\\ \hline
\\

\begin{tabular}{l}
$\mathrm{Fe}$  \\ 
BCC \\
$\mathrm{Im\bar{3}m}$\\
\end{tabular} & 
$-0.51 \pm 0.20$ &
-0.34 &
-0.48 &
-0.41 &
FM \footnotemark[1]&
-0.50 &
0.997&
0.994 \\

\\ 
\begin{tabular}{l}
Cr \\
BCC\\
$\mathrm{Im\bar{3}m}$\\
\end{tabular} &
$0.06\pm0.18$ & 
0.03&
0.06&
0.01&
AFM \footnotemark[3]&
0.06 &
0.630&
0.631\\

\\ 
\begin{tabular}{l}
Ni \\
FCC\\
$\mathrm{Fm\bar{3}m}$\\
\end{tabular} &
$-0.07\pm0.03$&
-0.06&
-0.06&
-0.06&
FM \footnotemark[3]&
-0.07&
0.998&
0.997\\

\\ 
\begin{tabular}{l}
$\mathrm{FePO_4}$ \\
$\alpha$-quartz\\
$\mathrm{P3_121}$\\
\end{tabular} &
$0.02\pm0.01$ &
0.02&
-0.01&
0.00&
AFM \footnotemark[4]&
0.02&
0.952&
0.962\\

\\ 
\begin{tabular}{l}
$\mathrm{FePO_4}$ \\
olivine\\
$\mathrm{Pnma}$\\
\end{tabular} &
$0.03\pm0.011$&
0.03&
0.02&
0.05&
AFM \footnotemark[5]&
0.03&
0.999&
0.998\\

\\ 
\begin{tabular}{l}
$\mathrm{LaMnO_3}$ \\
perovskite\\
$\mathrm{Pbnm}$\\
\end{tabular}&
$-0.07\pm0.02$&
-0.25&
-0.11&
-0.05&
FM \footnotemark[6]&
-0.07&
0.994&
0.998\\

\\ 
\begin{tabular}{l}
$\mathrm{CuCr_2O_4}$ \\
spinel\\
$\mathrm{I4_1/amd}$\\
\end{tabular}&
$-0.10\pm0.06$&
-0.07&
-0.11&
-0.05&
AFM1 \footnotemark[7]&
-0.10&
0.965&
0.969

\end{tabular}
\end{ruledtabular}
\footnotetext[1]{Ref.~\cite{Fe_exp_cell}}
\footnotetext[2]{Ref.~\cite{Chromium} This is a difference between nonmagentic (NM) and AFM}
\footnotetext[3]{Ref.~\cite{Cr_and_Ni_exp_cell}}
\footnotetext[4]{Ref.~\cite{FePO4_quartz}}
\footnotetext[5]{Ref.~\cite{FePO4_olivine_exp_cell}}
\footnotetext[6]{Ref.~\cite{LaMO3_high_T} The AFM state is A-type}
\footnotetext[7]{Ref.~\cite{CuCr2O4} This is a difference between two FiM states. See Figure \ref{fig:CuCr2O4}}
\end{table*}

The c-value of a particular magnetic ground state of a material may be used to understand when GGA-level DFT is giving a reliable prediction versus when a higher-order theory is needed. In the case of Fe, Ni, $\mathrm{FePO_4,~and~LaMnO_3}$ the c-values are larger than 0.9 indicating a nearly unanimous prediction of ferromagnetism in these materials. The high confidence is likely demonstrating the success of DFT in describing simple direct exchange ferromagnetism and indirect exchange antiferromagnetism as discussed earlier. However, the c-value is not always close to 1 as seen in the case of Cr. It is likely this number can be understood due to the fact that magnetic ground state of Cr is actually an incommensurate spin density wave as pointed out earlier and a proper understanding of Cr or any material with a spin density wave would require an extension of density functional theory as suggested by Capelle \textit{et. al}\cite{Capelle2000}. 

The utility of this c-value may not only lie in confirming when GGA has given you a prediction with high confidence, but it may also identify materials and material classes that have long been studied at the GGA level, but cannot be reliably understood due to the disagreement illuminated by the confidence value. By moving to more accurate, more computationally intensive methods, a truer understanding of these materials may lead to the discovery of new emergent phenomena.

\begin{table}
\caption{\label{tab:3}Calculated magnetic energy differences $E_{FM}-E_{AFM}$ in meV of various crystal structures using non-optimized experimental lattice parameters.}
\begin{ruledtabular}
\begin{tabular}{lcccccc}

Material & BEEF-vdW & PBE & RPBE & PBEsol & exp
\\ \hline
\\

\begin{tabular}{l}
$\mathrm{Fe}$  \\ 
BCC \\
$\mathrm{Im\bar{3}m}$\\
\end{tabular} & 
$-504.3 \pm 163.0$ &
-461.0 &
-477.9 &
-448.5 &
FM \\

\\ 
\begin{tabular}{l}
Cr \\
BCC\\
$\mathrm{Im\bar{3}m}$\\
\end{tabular} &
$63.6\pm165.4$ & 
42.6&
58.5&
28.5&
AFM \\

\\ 
\begin{tabular}{l}
Ni \\
FCC\\
$\mathrm{Fm\bar{3}m}$\\
\end{tabular} &
$-67.4\pm23.5$&
-63.5&
-65.9&
-62.0&
FM \\

\\ 
\begin{tabular}{l}
$\mathrm{FePO_4}$ \\
$\alpha$-quartz\\
$\mathrm{P3_121}$\\
\end{tabular} &
$27.1\pm7.0$ &
41.1&
35.4&
47.3&
AFM \\

\\ 
\begin{tabular}{l}
$\mathrm{FePO_4}$ \\
olivine\\
$\mathrm{Pnma}$\\
\end{tabular} &
$32.5\pm10.4$&
41.1&
35.4&
47.3&
AFM \\

\\ 
\begin{tabular}{l}
$\mathrm{LaMnO_3}$ \\
perovskite\\
$\mathrm{Pbnm}$\\
\end{tabular}&
$-66.322\pm27.1$&
-56.3&
-62.6&
-213.6&
FM \\

\\ 
\begin{tabular}{l}
$\mathrm{CuCr_2O_4}$ \\
spinel\\
$\mathrm{I4_1/amd}$\\
\end{tabular}&
$-100.4\pm70.5$&
-73.0&
-87.9&
-64.7&
AFM1 \footnotemark[1]

\end{tabular}
\end{ruledtabular}
\footnotetext[1]{This is a difference between two antiferromagnetic states. See Figure \ref{fig:CuCr2O4}}
\end{table}

\begin{table}
\caption{\label{tab:4} Comparison of the prediction confidence given by optimized and experiment lattice constants. The third column shows the reduction in energy attained from the DFT lattice optimization procedure rounded to the nearest 10 meV. The energy is given in eV.}
\begin{ruledtabular}
\begin{tabular}{lccc}
 & Optimized  & Experimental & Energy difference  
\\ \hline
\\
Fe&
0.997&
1.000&
-0.00\\

Cr&
0.630&
0.650&
-0.01\\

Ni&
0.998&
0.998&
-0.00\\

$\mathrm{FePO_4}$-q&
0.952&
1.000&
-0.01\\

$\mathrm{FePO_4}$-o&
0.999&
1.000&
-0.01\\

$\mathrm{LaMnO_3} $&
0.994&
0.994&
-0.01\\

$\mathrm{CuCr_2O_4}$&
0.965&
0.915&
-0.00
\end{tabular}
\end{ruledtabular}
\end{table}

\section{Case Study}
We demonstrate the application of our method to the case of two possible cathode materials. Density functional theory provides a simple way of calculating the theoretical voltage of an intercalation cathode with respect to a metal anode using the Nernst equation, $V=\mathrm{-\dfrac{\Delta G}{F}}$, where $\mathrm{\Delta G}$ is the Gibbs free energy per mole of the lithiation reaction and F is Faraday's constant. The voltage is written in this way so that a calculation of $\mathrm{\Delta G}$ per stoichiometric formula unit in eV directly relates to voltage in V. It has been shown previously that volume and entropic effects on the free energy change of lithium intercalation are on the scale of $10^{-5}$eV and $10^{-2}$eV respectively compared to the change in internal energy scale on the order of eV.\cite{Ceder1997} Therefore, it is common to use the change in internal energy at zero Kelvin to estimate the voltage. 

\begin{table*}
	\caption{\label{tab:5} We show the predicted magnetic ground state for optimized structures using BEEF-vdW along with PBE and PBE+U from Ramsan \textit{et. al}. We also present the c-value which incorporates all possible magnetic configurations.}
	\begin{ruledtabular}
		\begin{tabular}{lcccccccc}
			
			Material & BEEF-vdW  & PBE\footnotemark[1] & PBE+U\footnotemark[1] & c-value & \begin{tabular}{l} modified\\ c-value \end{tabular}
			\\ \hline
			\\
			
			$\mathrm{LiFePO_4F}$ & 
			AFM\footnotemark[2] &
			AFM&
			AFM&
			0.925 &
			0.993\\
			
			$\mathrm{NaFePO_4F}$ & 
			AFM\footnotemark[2]&
			AFM&
			AFM&
			0.567&
			0.991
			\\

		\end{tabular}
	\end{ruledtabular}
	\footnotetext[1]{Ref.~\cite{MagCathode1}}
	\footnotetext[2]{AFM interaction at all 3 length scales}
\end{table*}

Most Li-ion and Na-ion cathode materials typically exhibit magnetism in some form due to the presence of magnetic transition metal ions.  Previous works to identify the magnetic ground state of new Li-ion  and Na-ion cathode materials have shown inconsistent result for a few GGA-level functionals used. \cite{MagCathode1,MagCathode2} We explore a fluorinated iron phosphate cathode for both Li-ion and Na-ion and demonstrate the utility of our developed method to explore the magentic states of the material. The relevant reactions during operation are 

\[\mathrm{LiFePO_4F + Li^+ + e^- \rightleftharpoons Li_2 FePO_4F }\]
\[\mathrm{NaFePO_4F + Na^+ + e^- \rightleftharpoons Na_2 FePO_4F }\]

where the forward reaction represents discharge and the backward reaction represents charge.  Ramzan \textit{et. al}\cite{MagCathode1} have previously tested the FM and one of the possible AFM states of both the Na and Li-ion cathode materials above using the Perdew-Burke-Erzernhof (PBE) functional as well as the PBE-functional with a Hubbard correction of U=4.95 eV and J=0.95 eV (PBE+U). It is worth highlighting that there are multiple different lengths between two nearest Fe atoms which could lead to a collection of magnetic couplings either direct or mediated by oxygen, phosphorus, and/or fluorine. 
We explicitly consider 3 interaction lengths that can be spin aligned or anti-aligned. This leads to $2^3=8$ possible states. We therefore test the 7 possible AFM states as well as the FM state to properly predict the magnetic ground state.

Ramzan \textit{et. al} found a disagreement between PBE and PBE+U for the magnetic arrangement of both $\mathrm{Li_2FePO_4F}$ and $\mathrm{Na_2FePO_4F}$ but agreement that the ground state is AFM for $\mathrm{LiFePO_4F}$ and $\mathrm{NaFePO_4F}$. We were able to recreate these predictions of AFM ground states using the PBE as well as BEEF-vdW, but show a very low c-value of 0.567 for $\mathrm{Na_FePO_4F}$. The full results of magnetic prediction and c-values can be seen in Table \ref{tab:5}.  The example of $\mathrm{NaFePO_4F}$ clearly demonstrates the utility of the c-value as a robust and computationally efficient metric to identify consensus at the GGA level.

To further illustrate the importance of comparing multiple AFM  states, Table \ref{tab:5} includes a c-value encompassing all 8 calculated magnetic states as well as a modified c-value that only includes FM and the lowest energy AFM states, a comparison more like what is conventionally seen. The modified c-value is much higher than the more precise c-value, showing that there is relative certainty that the state is AFM but much less can be said about which specific AFM state is the ground state.

In the context of calculating the theoretical voltage for the intercalation cathode materials, the energy differences are typically small and thus, using the wrong magnetic state may not greatly affect the prediction. However, in other properties relevant to understanding the materials such as electronic conductivity, density of states, magnet moment for each Fe atom, and net magnetization, the correct magnetic state is vital. Only systematic searches, both in the possible magnetic states and spanning a collection of GGA functional, can give a reasonable expectation of getting these properties correct.

\section{Conclusion}
We presented a computationally efficient way to quantify uncertainty in the prediction of magnetic ground states of materials. We demonstrated the success of this method for various crystal structures, magnetic ordering and material classes. The method was then applied in the case of Li-ion and Na-ion cathode materials that contained magnetic transition metal ions. Our results show that in order to fully predict and understand the magnetic ground state of a material with GGA-level density functional theory, there must be a systematic sampling of both GGA functionals and of possible magnetic states. Our method provides a way to do both and therefore quantify relative certainty that one particular magnetic arrangement is more energetically favorable than the other. More importantly, it provides a simple framework with which to quote the certainty that there is a consensus at the GGA level for the predicted magnetic configurations.
\begin{acknowledgments}
	G. H. and V. V. acknowledge support from the Scott Institute for Energy Innovation at Carnegie Mellon.
\end{acknowledgments}

\bibliography{cite}

\end{document}